\documentstyle[12pt,aps,preprint,psfig]{revtex}

\begin{document}

\author{M. M. Islam \\
Physics Department, University of Connecticut,\\
Storrs, Connecticut, 06269, USA\ \\
e-mail: islam@main.phys.uconn.edu}
\title{Describing Hot and Dense Nuclear Matter with Gauged Linear $\sigma 
$-Model%
\thanks{$\,$ Talk given at the Fourth Workshop on Quantum Chromodynamics
(June, 1998). To be published in the Proceedings of the Workshop, ed. by
H.M. Fried and B. M\"{u}ller (World Scientific).}}
\date{\today }
\maketitle

\begin{abstract}
\ To describe nuclear matter at high temperature and high baryon density
appropriate for RHIC and LHC, an effective theory is proposed. Three
developments underlie the effective theory: (1) relativistic mean field
theory description of nuclear matter with mesons mediating interactions; (2)
topological soliton description of the nucleon with hidden local symmetry;
(3) phenomenological knowledge of nucleon-nucleon interaction and nucleon
structure obtained from elastic NN\ scattering at c.m.energies of hundreds
of GeV. When these developments are combined together, a gauged linear $%
\sigma $-model with anomalous action and condensed quark-antiquark ground
state emerges as the effective theory.
\end{abstract}

\thispagestyle{plain}

\section{Introduction}

A new and exciting frontier of physics is opening before us with the advent
of the Relativistic Heavy Ion Collider (RHIC) at Brookhaven and the Large
Hadron Collider (LHC) at CERN.$^{1}$ At RHIC, a 20 TeV gold nucleus will hit
head-on another 20 TeV gold nucleus, and the center of mass (c.m.) energy of
two colliding nucleons will be about 200 GeV. This is far from the energy
regime where nucleon-nucleon interaction has been extensively studied. Also,
at RHIC temperatures of the order of 200 MeV and nuclear densities about ten
times the normal nuclear density are expected to be reached. At LHC in
lead-lead collision, the c.m. energy of two colliding nucleons will be 5.4
TeV --- much greater than that at RHIC. Clearly, we have to confront
questions such as these:

1. What kind of theoretical framework do we envisage to describe this new
realm of physics, where nuclear matter reaches temperatures and densities
comparable to that of the early universe and the interior of neutron stars?

2. How do we formulate the anticipated phase transition in heavy-ion
collisions from the hadronic regime to the perturbative QCD regime?

Against this backdrop, I want to discuss an effective theory for describing
nuclear matter at high density and high temperature. This effective theory
emerges when the following three developments in three distinct but related
areas are combined together:

(1) Relativistic mean field theory (RMFT) based on quantum hadrodynamics
(QHD) models can describe realistically nuclear many-body effects and
properties of finite nuclei and can provide a suitable framework to study
the density and temperature dependence of nuclear matter. This conclusion
has been reached by Serot and Walecka after an extensive review of research
in this field.$^{2}$ We note that a QHD model is a field theory model based
on an effective local Lagrangian density, that has hadronic degrees of
freedom such as $\pi $, $\sigma $, $\omega $, $\rho $ and nucleons as basic
fermions.$^{3}$

(2) Extensive studies by various groups have shown that the nucleon can be
described as a topological soliton in the gauged nonlinear $\sigma $-model
(NL$\sigma $M) that has $\pi $, $\rho $, $\omega $ as the dynamical degrees
of freedom and that predicts successfully the low energy properties of the
nucleon down to distances of the order of 0.5 F.$^{4,5}$

(3) Phenomenological analysis of high energy elastic $pp$ and $\bar{p}p$
scattering in the c.m. energy range $\surd s=23-630$ GeV $^{6}$ shows strong
evidence in favor of the topological soliton model of the nucleon. The
analysis, however, also indicates that the soliton is embedded in a
condensed ground state of quarks and antiquarks.$^{7,8}$ The $q\bar{q}$
ground state is analogous to a superconducting ground state and provides an
outer cloud of the nucleon, which is responsible for diffraction scattering.

As we will see in the following sections, the effective theory we arrive at
is the gauged linear $\sigma $-model with hidden symmetry and spontaneous
breakdown of chiral symmetry. We note that effective theories --- based on
the Nambu-Jona-Lassinio model,$^{9}$ the Lee-Wick-Friedberg model,$^{10}$
the meson-quark-dilaton model $^{11}$ of the nucleon--- have been applied by
various groups to investigate the density and temperature dependence of
nuclear matter and chiral phase transition.$^{12-14}$ Our approach brings in
a new element to such investigations; namely, the phenomenological knowledge
of nucleon-nucleon interaction and nucleon structure obtained from the
analysis of high energy elastic scattering at NN c.m. energies of hundreds
of GeV --- an energy regime relevant to RHIC and LHC.

\section{A General Lagrangian Framework}

To see how the three developments mentioned in the Introduction can be
combined together, we first lay out a general Lagrangian framework. We begin
with the linear $\sigma $-model of Gell-Mann and Levy that has SU(2)$_{L}$ $%
\times $ SU(2)$_{R}$ global symmetry and spontaneous breakdown of chiral
symmetry. The model is given by the Lagrangian density 
\begin{eqnarray}
{\cal L} &=&\bar{\psi}\,i\,\gamma^{\mu }\,\partial_{\mu }\,\psi
+\frac{1}{2}\left( \partial_{\mu }\,\sigma \,\,\partial^{\mu }\,\sigma 
+\partial_{\mu }\,\vec{\pi}\,.\,\partial^{\mu }\,\vec{\pi}\right)   \nonumber \\
&&-G\,\bar{\psi}\,\left[ \sigma +i\,\vec{\tau}\,.\,\vec{\pi}\,\gamma
^{5}\right] \,\psi -\lambda \,
\left( \sigma^{2}+\vec{\pi}^{2}-f_{\pi}^{2} \right)^{2}.  \label{2.1}
\end{eqnarray}
We next introduce a scalar-isoscalar field $\zeta (x)$ and a unitary field $%
U(x)$ in the following way: 
\begin{equation}
\sigma (x)+i\,\vec{\tau}\,.\,\vec{\pi}(x)=\zeta (x)\,U(x).  \label{2.2}
\end{equation}
$\zeta (x)$ is the magnitude of the fields $\sigma (x)$ and $\vec{\pi}(x)$: $%
\zeta ^{2}(x)=\sigma ^{2}(x)+\vec{\pi}^{2}(x).$ $U(x)$ is given by $%
U(x)=\exp \left( i\,\vec{\tau}\,.\vec{\phi}(x)\,/\,f_{\pi }\right) $; here, $%
\vec{\phi}(x)$ is the isovector pion field, and $f_{\pi }$ is the pion decay
constant ($f_{\pi }\simeq 93$ MeV). The pions are the Goldstone bosons
generated by the spontaneous breakdown of SU(2)$_{A}$ symmetry. Let us next
introduce right and left fermion fields: 
\[
\psi _{R}=\frac{1}{2}\left( 1+\gamma ^{5}\right) \psi \text{,\thinspace
\thinspace \thinspace \thinspace \thinspace \thinspace \thinspace \thinspace
\thinspace \thinspace \thinspace \thinspace \thinspace \thinspace \thinspace
\thinspace \thinspace \thinspace \thinspace \thinspace \thinspace \thinspace
\thinspace \thinspace \thinspace }\psi _{L}=\frac{1}{2}\left( 1-\gamma
^{5}\right) \psi .
\]
In terms of these new fields, the Lagrangian density (\ref{2.1}) can be written 
as 
\begin{eqnarray}
{\cal L} &=&\bar{\psi}_{R}\,i\,\gamma ^{\mu }\,\partial _{\mu }\,\psi _{R}+%
\bar{\psi}_{L}\,i\,\gamma ^{\mu }\,\partial _{\mu }\,\psi _{L}+\frac{1}{2}%
\,\partial _{\mu }\,\zeta \,\,\partial ^{\mu }\,\zeta   \nonumber \\
&&+\frac{1}{4}\zeta ^{2}\,tr\left[ \partial _{\mu }\,U\,\,\partial ^{\mu
}\,U^{\dagger }\right] -G\,\zeta \,\left[ \bar{\psi}_{L}\,U\,\psi _{R}+\bar{%
\psi}_{R}\,U^{\dagger }\,\psi _{L}\right] -\lambda \,\left( \zeta
^{2}-f_{\pi }^{2}\right) ^{2}.  \label{2.3}
\end{eqnarray}
Under right and left global transformations: 
\begin{eqnarray}
\psi _{R} &\rightarrow &R\,\psi _{R}\,\,,\,\,\,\,\,\psi _{L}\rightarrow
L\,\psi _{L}\,;  \label{2.4a} \\
U(x) &\rightarrow &L\,U(x)\,R^{\dagger }.  \label{2.4b}
\end{eqnarray}
Eq.(\ref{2.3}) makes it evident that the Lagrangian is invariant under these
transformations and that the scalar-isoscalar field $\zeta (x)$ is chiral
invariant. When $\zeta (x)$ is replaced by its vacuum value $f_{\pi }$, the
Lagrangian (\ref{2.3}) represents a nonlinear $\sigma $-model (NL$\sigma $M).

Following Bando et al., $^{15}$ we now introduce the idea of a hidden local
symmetry. Bando et al. took this symmetry to be [SU(2)]$_{\text{hidden}}$.
We consider an extended version of this approach, and following Meissner et
al. $^{16}$ take the symmetry to be [SU(2)$\times $U(1)]$_{\text{hidden}}$.
To implement the idea of a hidden local symmetry, we write $U(x)=\xi
_{L}^{\dagger }(x)\,\xi _{R}(x)$, where $\xi _{R}(x)$, $\xi _{L}(x)$ are
SU(2)-valued fields which transform in the following way under [SU(2)$%
_{L}\times $SU(2)$_{R}$]$_{\text{global}}$ $\times $ [SU(2)$\times $U(1)]$_{%
\text{local}}:$%
\begin{equation}
\xi _{R}(x)\rightarrow h(x)\,\xi _{R}(x)\,R^{\dagger }\text{,\thinspace
\thinspace \thinspace \thinspace \thinspace \thinspace \thinspace \thinspace
\thinspace \thinspace \thinspace \thinspace \thinspace \thinspace \thinspace
\thinspace \thinspace \thinspace \thinspace \thinspace \thinspace \thinspace
\thinspace \thinspace \thinspace }\xi _{L}(x)\rightarrow h(x)\,\xi
_{L}(x)\,L^{\dagger },  \label{2.5}
\end{equation}
where $h(x)\,\epsilon \,$[SU(2)$\times $U(1)]$_{\text{local}}$. Eq(\ref{2.5})
shows that 
\begin{equation}
U(x)=\xi _{L}^{\dagger }(x)\,\xi _{R}(x)\rightarrow L\,U(x)\,R^{\dagger } 
\label{2.6}
\end{equation}
as required, so that the global symmetry is maintained.

Let us focus on the pion sector of the Lagrangian (\ref{2.3}): 
\begin{eqnarray}
{\cal L}_{\pi } &=&\frac{1}{4}\zeta ^{2}\,tr\left[ \partial _{\mu
}\,U\,\,\partial ^{\mu }\,U^{\dagger }\right]  \nonumber \\
&=&-\frac{1}{4}\zeta ^{2}\,tr\left[ \partial _{\mu }\,\xi _{L}\,\xi
_{L}^{\dagger }-\partial _{\mu }\,\xi _{R}\,\xi _{R}^{\dagger }\right] ^{2}.
\label{2.7}
\end{eqnarray}
We gauge this symmetry by introducing the vector mesons $\vec{\rho}$ and $%
\omega $ as gauge bosons of the [SU(2)$\times $U(1)]$_{\text{hidden}}$
symmetry. Gauging is done easily by replacing the ordinary derivative by the
covariant derivative: $\partial _{\mu }\rightarrow D_{\mu }=\partial _{\mu
}+V_{\mu }$, where 
\begin{equation}
V_{\mu }=-i\,\frac{g}{2}\,\left[ \vec{\tau}\,.\,\vec{\rho}_{\mu }+\omega
_{\mu }\right] ,  \label{2.8}
\end{equation}
and $V_{\mu }$ transforms under the hidden symmetry in the following way:

\begin{equation}
V_{\mu }\rightarrow h\,V_{\mu }\,h^{\dagger }+h\,\partial _{\mu
}\,h^{\dagger }.  \label{2.9}
\end{equation}

The Lagrangian density of the gauged hidden-symmetry model is taken as $%
^{15,16}$ 
\begin{eqnarray}
{\cal L}_{\pi } &=&-\frac{1}{4}\zeta ^{2}\,tr\left[ \left( D_{\mu }\,\xi
_{L}\right) \,\xi _{L}^{\dagger }-\left( D_{\mu }\,\xi _{R}\right) \,\xi
_{R}^{\dagger }\right] ^{2}  \nonumber \\
&&-\frac{1}{2}\zeta ^{2}\,tr\left[ \left( D_{\mu }\,\xi _{L}\right) \,\xi
_{L}^{\dagger }+\left( D_{\mu }\,\xi _{R}\right) \,\xi _{R}^{\dagger
}\right] ^{2}+\frac{1}{2g^{2}}tr\left[ F_{\mu \nu }\,F^{\mu \nu }\right] , 
\label{2.10}
\end{eqnarray}
where the first term is exactly the same as the original term $\frac{1}{4}%
\zeta ^{2}\,tr\left[ \partial _{\mu }\,U\,\,\partial ^{\mu }\,U^{\dagger
}\right] .$ The second term is a gauge invariant term that generates the
masses of the vector mesons. The third term is the Lagrangian density of the
gauge field $V_{\mu }(x)$, and $F_{\mu \nu }$ is the nonabelian field
tensor: $F_{\mu \nu }=\partial _{\mu }\,V_{\nu }-\partial _{\nu }\,V_{\mu
}+\left[ V_{\mu },V_{\nu }\right] .$ The last term in (\ref{2.10}) produces a 
term
like $tr\left[ L_{\mu },L\nu \right] ^{2}$, where $L_{\mu }=U^{\dagger
}\,\partial _{\mu }\,U.^{4}$ This is the fourth-order derivative term
introduced by Skyrme.

We next focus on the fermion-scalar-meson sector of the Lagrangian density
(\ref{2.3}) and introduce new fermionic variables: 
\begin{equation}
\psi _{R}^{0}(x)=\xi _{R}(x)\,\psi _{R}(x)\text{,\thinspace \thinspace
\thinspace \thinspace \thinspace \thinspace \thinspace \thinspace \thinspace
\thinspace \thinspace \thinspace \thinspace \thinspace \thinspace \thinspace
\thinspace \thinspace \thinspace \thinspace \thinspace \thinspace \thinspace
\thinspace \thinspace }\psi _{L}^{0}(x)=\xi _{L}(x)\,\psi _{L}(x). 
\label{2.11}
\end{equation}
These variables are invariant under the global right and left
transformations. They transform only under the local hidden symmetry: 
\begin{equation}
\psi _{R}^{0}(x)\rightarrow h(x)\,\psi _{R}^{0}(x)\text{, }\psi
_{L}^{0}(x)\rightarrow h(x)\,\psi _{L}^{0}(x)  \label{2.12}
\end{equation}

\[
(h(x)\,\epsilon \,[SU(2)\times U(1)]_{\text{hidden}}) 
\]

In terms of these new variables, the fermion-scalar-meson sector of (\ref{2.3})
becomes 
\begin{eqnarray}
{\cal L}_{F+S} &=&\bar{\psi}^{0}\,i\,\gamma ^{\mu }\left[ \partial _{\mu
}+V_{\mu }+\tilde{v}_{\mu }+\tilde{a}_{\mu \,}\,\gamma ^{5}\right] \psi ^{0}
\nonumber \\
&&+\frac{1}{2}\partial _{\mu }\,\zeta \,\partial ^{\mu }\,\zeta -G\,\zeta \,%
\bar{\psi}^{0}\,\psi ^{0}-\lambda \,\left( \zeta ^{2}-f_{\pi }^{2}\right)
^{2},  \label{2.13}
\end{eqnarray}
where 
\begin{equation}
\psi ^{0}=\psi _{R}^{0}+\psi _{L}^{0}  \label{2.14a}
\end{equation}
\begin{equation}
\tilde{v}_{\mu }=\frac{1}{2}\left[ \xi _{R}\,\partial _{\mu }\,\xi
_{R}^{\dagger }+\xi _{L}\,\partial _{\mu }\,\xi _{L}^{\dagger }\right]
-V_{\mu },  \label{2.14b}
\end{equation}
\begin{equation}
\tilde{a}_{\mu }=\frac{1}{2}\left[ \xi _{R}\,\partial _{\mu }\,\xi
_{R}^{\dagger }-\xi _{L}\,\partial _{\mu }\,\xi _{L}^{\dagger }\right] . 
\label{2.14c}
\end{equation}
Both $\tilde{v}_{\mu }$ and $\tilde{a}_{\mu }$ transform covariantly under
the hidden symmetry: 
\[
\tilde{v}_{\mu }(x)\rightarrow h(x)\,\tilde{v}_{\mu }(x)\,h^{\dagger }(x),
\]
\[
\tilde{a}_{\mu }(x)\rightarrow h(x)\,\tilde{a}_{\mu }(x)\,h^{\dagger }(x).
\]
Since the covariant derivative also transforms covariantly, 
\[
\partial _{\mu }+V_{\mu }\rightarrow h\,(\partial _{\mu }+V_{\mu
})\,h^{\dagger },
\]
the Lagrangian density ${\cal L}_{F+S}$ is manifestly invariant under the
hidden gauge symmetry, besides being invariant under the global symmetry.

\section{Emergence of an Effective Theory}

We observe that the Lagrangian density in Eq.(\ref{2.13}) is very similar to the
Lagrangian density of the nucleon-scalar-meson sector of QHD-II, which has $%
\omega $, $\rho $, $\zeta $ (scalar-isoscalar meson) as the hadronic degrees
of freedom besides $\pi .^{2,3}$ This similarity, of course, suggests that
we identify $\psi ^{0}$ as the nucleon field and allow the potential energy
density of the scalar field ${\cal V}(\zeta )=\lambda \,\left( \zeta
^{2}-f_{\pi }^{2}\right) ^{2}$ to be of more general form: ${\cal V}(\zeta
)=a\,\zeta ^{2}+b\,\zeta ^{3}+c\,\zeta ^{4}$, so as to fit the bulk
properties of finite nuclei and equilibrium properties of nuclear matter.
Furthermore, if we take $\xi _{L}^{\dagger }=\xi _{R}=\xi $, so that $U=\xi
^{2}$, then under the hidden local transformation: 
\begin{equation}
\psi _{R}^{0}(x)\rightarrow h(x)\,\psi _{R}^{0}(x)\text{,}%
\,\,\,\,\,\,\,\,\,\,\,\,\,\,\,\,\,\,\,\,\,\,\psi _{L}^{0}(x)\rightarrow
h(x)\,\psi _{L}^{0}(x),  \label{2.15a}
\end{equation}
and $h(x)$ has to satisfy the requirement 
\begin{equation}
L\,\xi (x)\,h^{\dagger }(x)=h(x)\,\xi (x)\,R^{\dagger }\text{,}  \label{2.15b}
\end{equation}
i.e., the requirement for nonlinear realization of the chiral symmetry.$^{2}$
Eqs.(\ref{2.15a},\ref{2.15b}) are, indeed, the requirements imposed on the nucleon field
both in QHD-II and in its extended version that incorporates the trace
anomaly.$^{17}$ We also note that the scalar field $\zeta $ in (\ref{2.13}), 
which
has a chiral invariant interaction $G\,\zeta \,\bar{\psi}^{0}\,\psi ^{0}$,
can be identified with the scalar-isoscalar field of QHD. It is not the $%
\sigma $-field of single-boson exchange potentials with $m_{\sigma }\simeq
500$ MeV. In fact, if we write $\zeta (x)=f_{\pi }+\hat{\zeta}(x)$, then the
fluctuation field $\hat{\zeta}(x)$ is to be identified with the $\sigma $%
-field. As Kalafatis and Vinh Mau have shown,$^{18}$ it is the fluctuation
field $\hat{\zeta}(x)$ that provides the NN medium range attraction. It is
evident that the Lagrangian (\ref{2.13}) with a generalized ${\cal V}(\zeta )$
incorporates all the main features of nucleon-scalar -meson sector of QHD-II.

Turning to the Lagrangian density ${\cal L}_{\pi }$ in Eq.(\ref{2.10}), let us
consider that the scalar field $\zeta (x)$ can be replaced by its vacuum
value $f_{\pi }$. Following Meissner et al.,$^{16}$ we add to it the
anomalous action, i.e., the Wess-Zumino-Witten (WZW) action: 
\begin{equation}
{\cal L}_{WZW}=g_{\omega }\,\omega _{\mu }\,B^{\mu },  \label{2.16}
\end{equation}
where $B_{\mu }$ is the topological baryonic current: 
\begin{equation}
B^{\mu }=\frac{1}{24\,\pi ^{2}}\epsilon ^{\mu \nu \rho \sigma }\,tr\left[
U^{\dagger }\,\partial _{\nu }\,U\,U^{\dagger }\,\partial _{\rho
}\,U\,U^{\dagger }\,\partial _{\sigma }\;U\right] ,  \label{2.17}
\end{equation}
and $U=\exp \left[ i\,\vec{\tau}.\vec{\phi}\,/\,f_{\pi }\right] $. The
Lagrangian density we have now: ${\cal L}_{\pi }+{\cal L}_{WZW}$ represents
the gauged nonlinear $\sigma $-model with its anomalous action. This model
has been extensively studied by many groups.$^{4,5,19}$ It describes the
nucleon as a topological soliton and predicts successfully the low energy
properties of the nucleon. Also, the masses of $\rho $ and $\omega $ are
generated dynamically from the gauge-invariant second term in Eq.(\ref{2.10}) 
and
leads to the KSFR relation $m_{\rho }^{2}=m_{\omega }^{2}=2\,f_{\pi
}^{2}\,g^{2}.$ The last term in (\ref{2.10}), as we noted earlier, reproduces
Skyrme's fourth order derivative term.

The discussion above shows that the Lagrangian density 
\begin{equation}
{\cal L}={\cal L}_{\pi }+{\cal L}_{F+S}+{\cal L}_{WZW}  \label{2.18}
\end{equation}
with a generalized ${\cal V}(\zeta )$ encompasses both QHD-II model for
nuclear matter and NL$\sigma $M for the nucleon. If, in the above
Lagrangian, we take the fermion to be the nucleon, drop ${\cal L}_{WZW}$ and
replace $\zeta (x)$ in ${\cal L}_{\pi }$ by its vacuum value $f_{\pi }$,
then we essentially obtain the QHD-II\ model.$^{2}$ On the other hand, if we
keep ${\cal L}_{\pi }$ with $\zeta (x)=f_{\pi }$ and ${\cal L}_{WZW}$, but
drop ${\cal L}_{F+S}$, then we have the gauged NL$\sigma $M.$^{16,19}$ We now
examine how analysis of high energy elastic scattering impacts on these two
developments.

High energy elastic scattering at the CERN ISR and SPS Collider in the c.m.
energy range $\surd s=23-630$ GeV has been analyzed by my collaborators and
me over a number of years.$^{6}$ From this analysis, we arrived at the
following phenomenological description. The nucleon has a core and an outer
cloud. High energy elastic scattering is primarily due to two processes: 1)
a glancing collision where the outer cloud of a nucleon interacts with that
of the other and gives rise to diffraction scattering; 2) a hard collision
where one nucleon core scatters off the other core via $\omega $ exchange,
while their outer clouds overlap and interact independently. In the small
momentum transfer region diffraction dominates, but as the momentum transfer
increases, the hard scattering takes over.$^{8}$

Clearly, elastic scattering shows that the nucleon is a composite object
with a core and a cloud. Hence, nucleons cannot be regarded as basic
fermions as in QHD-II at c.m. energies of hundreds of GeV. Instead, the
basic fermions should be taken as more fundamental objects---quarks
(perhaps, more precisely, effective quarks). Also, a change of fermionic
variables is carried out in QHD-II to implement the nonlinear realization of
chiral symmetry (Eq.(\ref{2.11})). Path integral formalism shows that, because 
of
the fermion measure, change of fermionic variables induces an anomalous
action,$^{20}$ which in this case is the WZW\ action. Therefore, the WZW
action has to be included in the effective model. However, as we discuss
below, the dynamics of elastic scattering and nucleon structure indicate
that $\pi $, $\zeta $, $\omega $ are appropriate effective degrees of
freedom in the high energy region just as they are in the low energy region
in QHD-II. Furthermore, the nonlinear realization of chiral symmetry
implemented in the Lagrangian (\ref{2.18}) has also been found to be more
appropriate to describe properties of finite nuclei, such as charge density,
spin-orbit splitting, etc.$^{17}$ Hence, a relativistic mean field theory
based on the Lagrangian (\ref{2.18}) with quarks as the basic fermions should be 
a
realistic framework for nuclear matter.

With regard to the gauged NL$\sigma $M, elastic scattering analysis supports
the conclusion that the nucleon is a topological soliton. In Fig.1, the pion
profile function calculated from high energy elastic scattering$^{8}$ is
compared with the pion profile functions obtained by Meissner et al.$^{19}$
from low energy considerations. The profile functions are quite compatible
even though they come from two totally different domains of physics. The
gauged NL$\sigma $M, however, completely ignores the scalar field $\zeta $
---replacing it from the very beginning by its vacuum value $f_{\pi }$ and
neglecting the important interaction that $\zeta $ mediates between left and
right quarks. The only contribution NL$\sigma $M keeps from the fermion
sector is the anomalous action arising from the gauge dependence of the
fermion measure. This implies that the soliton of the NL$\sigma $M lies in a
noninteracting Dirac sea (Fig. 2a). If, on the other hand, we keep the
scalar field $\zeta $, assume that replacing it by its vacuum value $f_{\pi }
$ is reasonable in the pion sector, then in the fermion sector the left and
right quarks still interact via the scalar field. In this case, we have a
soliton that lies in an interacting Dirac sea (Fig. 2b). What one finds is
that, if this scalar field has a critical behavior (by this I mean it is
zero at small distances, but rises sharply at some distance $r=R$ to its
vacuum value $f_{\pi }$ as in Fig. 2c), then the interacting Dirac sea can
have considerably less energy than the noninteracting Dirac sea.$^{7}$ The
system, in this case, makes a phase transition to the interacting ground
state and reduces its total energy substantially by the condensation energy.
The phenomenon is analogous to superconductivity. It also solves a
persistent problem of the NL$\sigma $M; namely, its prediction of a large
soliton mass ($\sim 1500$ MeV) compared to the actual mass (939 MeV) of the
nucleon. Such a condensation phenomenon implies that the nucleon is a
topological soliton embedded in a $q\bar{q}$ ground state, and the ground
state provides an outer cloud. High energy elastic scattering analysis, as
mentioned earlier, shows evidence of such a cloud that gives rise to
diffraction scattering. A number of other consequences also follow from
having a condensed $q\bar{q}$ ground state.$^{8}$

\section{Summary and Conclusion}

We have discussed three independent but related developments:

(1) Relativistic mean field theory description of nuclear matter based on a
QHD model with $\pi $, $\omega $, $\rho $, and $\zeta $ (scalar-isoscalar
meson) as the hadronic degrees of freedom.

(2) Topological soliton description of the nucleon based on a gauged
nonlinear $\sigma $-model with hidden local symmetry and anomalous action.

(3) Phenomenological knowledge of nucleon-nucleon interaction and nucleon
structure obtained from analysis of high energy elastic scattering in the
c.m. energy range $\surd s=23-630$ GeV.

These three developments when combined together strongly suggest that the
Lagrangian density 
\setcounter{equation}{21}
\begin{equation}
{\cal L}={\cal L}_{\pi }+{\cal L}_{F+S}+{\cal L}_{WZW} 
\end{equation}
should provide a very realistic effective theory to investigate density and
temperature dependence of nuclear matter over a wide range. Fermions in this
effective theory are to be taken as ``quarks'' and not as nucleons.
Furthermore, potential energy density ${\cal V}(\zeta )$ of the scalar field
should be taken to have a form more general than the Higgs potential in
(\ref{2.3}). This can lead to a physical vacuum at $\zeta =f_{\pi }$ and a
metastable or false vacuum at $\zeta =0$.$^{13}$ Since in mean field theory,
meson fields are replaced by their constant vacuum expectation values, and
only the fermion field is quantized, it is straightforward to obtain the
thermodynamic potential $\Omega $ from the grand partition function.
Quantities such as energy density, baryon density, and pressure can then be
obtained from $\Omega $, and meson field eqautions can also be derived by
extremizing it.$^{3,14}$

Investigation of high energy elastic scattering dynamics has shown that the
scalar-isoscalar field $\zeta $, which is an essential element of quantum
hadrodynamics models, plays the crucial role of an order parameter. Its
critical behavior leads to a $q\bar{q}$ condensed ground state analogous to
a superconducting ground state. Furthermore, vector meson $\omega $ behaves
as a gauge boson coupled to the topological baryonic charge as in the gauged
nonlinear $\sigma $-model with WZW action. Obviously, the fields $\pi $, $%
\zeta $, and $\omega $, which are appropriate degrees of freedom in the low
energy region, also constitute appropriate degrees of freedom at NN c.m.
energies of hundreds of GeV. Hence, a relativistic mean field theory based
on the Lagrangian density (\ref{2.18}) that has these degrees of freedom and
quarks as basic fermions should provide a realistic framework to study
nuclear matter at high densities and high temperatures---densities and
temperatures that will be reached at RHIC and LHC.\vspace{0.5cm}

The author wishes to thank Brian Serot and Robert Vinh Mau for their
comments. \vspace{1cm}

\noindent {\Large {\bf References}} 

\begin{itemize}
\begin{description}
\item  \noindent $1.\,$J.W. Harris and B. M\"{u}ller, Annu. Rev. Nucl. Part.
Sci. \underline{46}, 71 (1996).

\item  $2.\,$B.D. Serot and J.D. Walecka, Int. J. of Mod. Phys. E \underline{%
6}, 515 (1997).

\item  $3.\,$J.D. Walecka, Theoretical Nuclear and Subnuclear Physics
(Oxford Univ. Press, 1995).

\item  $4.\,$R.K. Bhaduri, Models of the Nucleon: From Quarks to Soliton
(Addison-Wesley Publishing Co., 1988).

\item  $5.\,$U.G. Meissner, Phys. Rep. \underline{161}, 213 (1988).

\item  $6.\,$M.M Islam, V. Innocente, T. Fearnley and G. Sanguinetti,
Europhys. Lett. \underline{4}{\bf , }189 (1987){\bf \ }and references
therein.

\item  $7.\,$M.M Islam, Z. Phys. C \underline{53}, 253 (1992).

\item  $8.\,$M.M. Islam, in Proceedings of the Quantum Infrared Workshop,
ed. by H.M. Fried and B. M\"{u}ller (World Scientific, 1995) p. 401.

\item  $9.\,$R. Alkofer, H. Reinhardt, and H. Weigel, Phys. Rep. \underline{%
265}, 139 (1996).

\item  $10.\,$T.D. Lee and Y.Pang, Phys. Rep. \underline{221}{\bf , }251
(1992).

\item  $11.\,$P. Jain, R. Johnson, and J. Schechter, Phys. Rev. D \underline{%
38}, 1571 (1988).

\item  $12.\,$T. Hatsuda and T. Kunihiro, Phys. Rep. \underline{247}, 221
(1994).

\item  $13.\,$W.N. Cottingham and R. Vinh Mau, Phys. Rev. D \underline{44},
1652 (1991); W.N. Cottingham, D. Kalafatis, and R. Vinh Mau, Phys. Rev.
Lett. \underline{73}, 1328 (1994).

\item  $14.\,$P. Papazoglou, J. Schaffner, S. Schramm, D. Zschiesche, H.
St\"{o}cker, and W. Greiner, Phys. Rev. C \underline{55}, 1499 (1997).

\item  $15.\,$M. Bando, T. Kugo, S. Uehara, K. Yamawaki, and T. Yanagida,
Phys. Rev. Lett. \underline{54}, 1215 (1985).

\item  $16.\,$U. G. Meissner, N. Kaiser. A. Wirzba, and W. Weise, Phys. Rev.
Lett. \underline{57}, 1676 (1986).

\item  $17.\,$R.J. Furnstahl, H.B. Tang, and B.D. Serot, Phys. Rev. C 
\underline{52}, 1368 (1995).

\item  $18.\,$D. Kalafatis and R. Vinh Mau, Phys. Lett. B \underline{283},
13 (1992).

\item  $19.\,$U.G. Meissner, N. Kaiser, and W. Weise, Nucl. Phys. A 
\underline{466}, 685 (1987).

\item  $20.\,$M.M. Islam and S.J. Puglia, Int. J. of Mod. Phys. A \underline{%
13}, 523 (1998).
\end{description}
\end{itemize}
\begin{minipage}{150mm}
\begin{center}
\mbox{\psfig{file=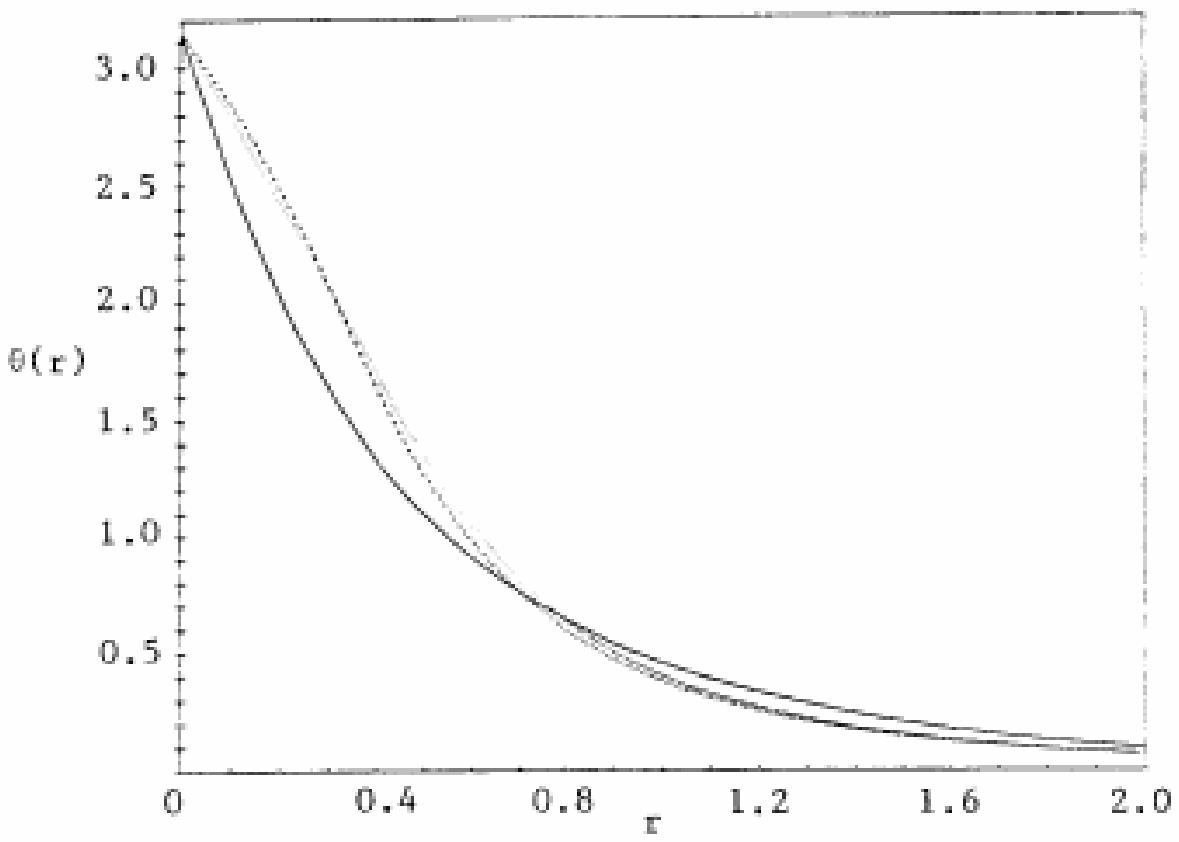}}
\end{center}
Fig.1. The pion profile function $\theta(r)$ as a function of $r$ in
fermi. The continuous curve represents $\theta(r)$ obtained from high energy
elastic scattering. The dotted and the dashed curves represent $\theta(r)$ 
calculated from low energy in the minimal and in the complete soliton model 
by Meissner et. al. [Ref.19]
\vfill
\begin{center}
\mbox{\psfig{file=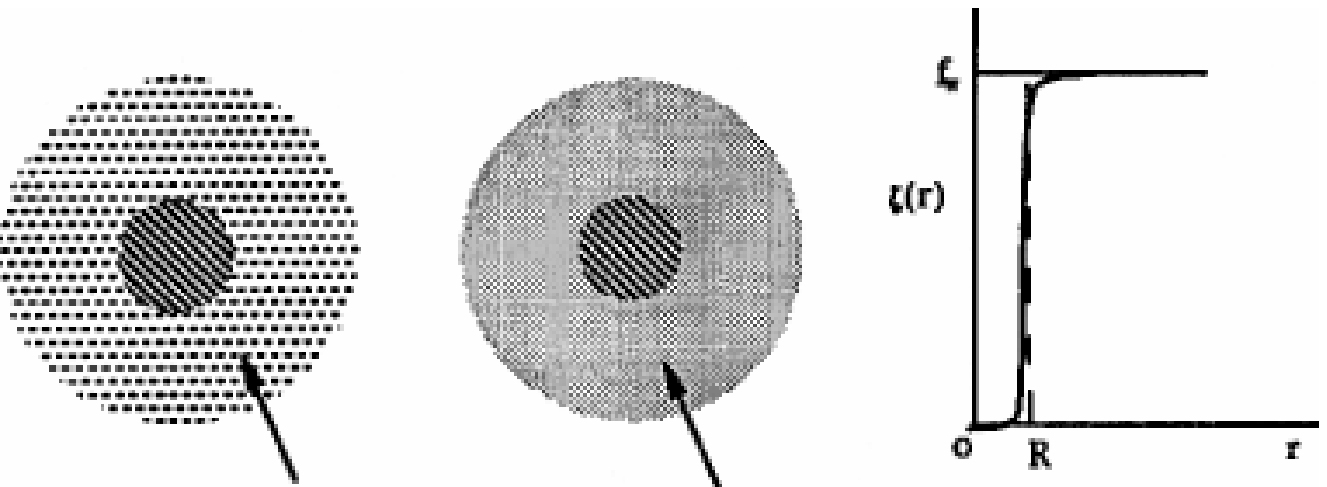,width=5.4in}}
\end{center}
\begin{center}
 Noninteracting Dirac sea\hspace*{7mm}Interacting Dirac sea
\hspace*{3mm}Critical behavior of $\zeta(r)$ \\
Fig.2a.\hspace*{3.4cm}Fig.2b.\hspace*{3.3cm}Fig.2c.\\
\end{center}
\end{minipage}

\end{document}